\documentstyle[11pt]{article}

\begin{document}
\large

\def\lsim{\mathrel{\rlap{\lower3pt\hbox{\hskip0pt$\sim$}}
    \raise1pt\hbox{$<$}}}         
\def\gsim{\mathrel{\rlap{\lower4pt\hbox{\hskip1pt$\sim$}}
    \raise1pt\hbox{$>$}}}         
\def\dblint{\mathop{\rlap{\hbox{$\displaystyle\!\int\!\!\!\!\!\int$}}
    \hbox{$\bigcirc$}}}
\def\ut#1{$\underline{\smash{\vphantom{y}\hbox{#1}}}$}

\newcommand{\beq}{\begin{equation}}
\newcommand{\eeq}{\end{equation}}
\newcommand{\dem}{\Delta M_{\mbox{B-M}}}
\newcommand{\dega}{\Delta \Gamma_{\mbox{B-M}}}

\newcommand{\ind}[1]{_{\begin{small}\mbox{#1}\end{small}}}
\newcommand{\WA}{{\em WA}}
\newcommand{\SM}{Standard Model }
\newcommand{\QCD}{{\em QCD }}
\newcommand{\KM}{{\em KM }}

\newcommand{\appa}{\mbox{\ae}}
\newcommand{\CP}{{\em CP}}
\newcommand{\fy}{\varphi}
\newcommand{\hi}{\chi}
\newcommand{\al}{\alpha}
\newcommand{\as}{\alpha_s}
\newcommand{\gf}{\gamma_5}
\newcommand{\de}{\delta}
\renewcommand{\Im}{\mbox{Im}}
\renewcommand{\Re}{\mbox{Re}}
\newcommand{\GeV}{\mbox{ GeV }}
\newcommand{\MeV}{\mbox{ MeV }}
\newcommand{\matel}[3]{\langle #1|#2|#3\rangle}

\begin{center}\LARGE{Theoretical Physics Institute
\\University of Minnesota}
\end{center}
\begin{flushright}
\large{
TPI-MINN-92/67-T
\\ UND-HEP-92-BIG\hspace{0.1em}07
\\ NSF-ITP-92-156}
\\ December 1992
\end{flushright}
\vspace{0.2cm}
\begin{center} \LARGE {
A QCD    \lq\lq Manifesto\rq\rq on Inclusive Decays of Beauty and
Charm\footnote{talk presented at DPF meeting of APS, November 1992}}
\end{center} \vspace{0.2cm}
\begin{center} \Large I.I.Bigi
\\{\normalsize\it Dept.of Physics,
University of Notre Dame du
Lac, Notre Dame, IN 46556} \vspace{0.2cm}
\\B.Blok
\\{\normalsize\it Institute for Theoretical Physics, University of
California at Santa Barbara, Santa Barbara, CA 93106} \vspace{.2cm}
\\M.Shifman
\\{\normalsize\it Theoretical Physics Institute,
University of Minnesota, Minneapolis, MN 55455} \vspace{.2cm}
\\ N.G.Uraltsev
\\{\normalsize\it Dept.of
Physics,
University of Notre Dame du Lac, Notre Dame, IN 46556
\\{\normalsize and}\\
St.Petersburg Nuclear Physics Institute,
Gatchina, St.Petersburg 188350, Russia\footnote{permanent address}}
\vspace{0.2cm}
\\ A.Vainshtein
\\{\normalsize\it Theoretical Physics Institute,
University of Minnesota, Minneapolis, MN 55455
\\{\normalsize and}\\
Budker Institute for Nuclear Physics,
Novosibirsk 630090, Russia}
\end{center}
\thispagestyle{empty} \vspace{.4cm}

\centerline{\Large\bf Abstract}
\vspace{.3cm}
\noindent
A selfconsistent treatment of nonperturbative effects in inclusive nonleptonic
and semileptonic decays of beauty and charm hadrons is presented. It is
illustrated by calculating semileptonic branching ratios, lifetime ratios,
radiative decay rates and the lepton spectra in semileptonic decays.

\newpage
\font\tenbf=cmbx10
\font\tenrm=cmr10
\font\tenit=cmti10
\font\elevenbf=cmbx10 scaled\magstep 1
\font\elevenrm=cmr10 scaled\magstep 1
\font\elevenit=cmti10 scaled\magstep 1
\font\ninebf=cmbx9
\font\ninerm=cmr9
\font\nineit=cmti9
\font\eightbf=cmbx8
\font\eightrm=cmr8
\font\eightit=cmti8
\font\sevenrm=cmr7

\newcommand{\la}{\langle}
\newcommand{\ra}{\rangle}
\renewcommand{\Im}{\,\mbox{Im}\:}

\textwidth 6.0in
\textheight 8.5in
\topmargin -0.25truein
\newcommand{\bibit}{\nineit}
\newcommand{\bibbf}{\ninebf}
\renewenvironment{thebibliography}[1]
 { \elevenrm
   \begin{list}{\arabic{enumi}.}
    {\usecounter{enumi} \setlength{\parsep}{0pt}
     \setlength{\itemsep}{3pt} \settowidth{\labelwidth}{#1.}
     \sloppy
    }}{\end{list}}

\parindent=3pc
\baselineskip=10pt

{\elevenbf\noindent 1. Sketch Of General Method}
\vglue 0.2cm
\baselineskip=14pt
\elevenrm
It is the goal of our analysis to calculate transition rates for inclusive
nonleptonic and semileptonic decays in terms of the fundamental
parameters, namely quark masses, KM parameters and $\Lambda _{QCD}$ without
invoking Voodoo. Nonperturbative effects -- namely boundstate dynamics in the
initial state and hadronization in the final state -- have to be incorporated
here. We will do that via an expansion in $1/m_Q$ ($\,m_Q$ is the mass of the
heavy
flavour quark $Q$). For the beauty quark
mass is certainly much larger than
typical hadronic scales, and an extrapolation down to charm hopefully still
makes sense.

The width for the decay of a
hadron with the heavy flavor $Q$ into an inclusive
final state $X$ is obtained from the transition operator
$\hat T (Q\rightarrow X \rightarrow Q)$ describing the forward scattering of a
heavy quark $Q$ via an intermediate state $X$. Then we take the imaginary
part of the matrix element of
$\hat T$ between the state $H_Q$:
$$ \Gamma (H_Q \rightarrow X)\; \;    \propto \; \;G_F^2\;
\la H_Q|\Im \hat T(Q\rightarrow X\rightarrow Q)|H_Q\ra . \eqno(1)$$
One actually proceeds in two steps to evaluate this quantity:

\noindent {\elevenbf (a)} The nonlocal second-order
operator $\hat T$ is expanded into a series of local
operators $O^{(i)}$ of increasing dimension with coefficients $c_i$ that
contain increasing powers of $1/m_Q$
$$ \Im\, \hat T(Q\rightarrow X\rightarrow Q)\simeq \sum _i c_i O^{(i)}
\eqno(2)$$
(Strictly speaking, the expansion parameter is $1/E$ where $E$ is the energy
release, rather than $1/m_Q$.)
The lowest dimensional operators $O^{(i)}$ thus dominate for
large $m_Q$.

\noindent {\elevenbf (b)} One has to evaluate the matrix elements
$\la H_Q|O^{(i)}|H_Q\ra $; it is at
this level that one differentiates between the decays of heavy flavor
baryons and neutral and charged pseudoscalar mesons, i.e. $\Lambda _Q$, $P^0$
and $P^{\pm}$. This approach was explicitely formulated in ref.$^1$ and
has been supported by the analysis of ref.$^2$.

(1) The general expansion of $\Im \, \hat T$ is as follows (for details
see $^{1-4}$)
$$ \Im\, \hat T(Q\rightarrow X\rightarrow Q)=
\tilde c^{(X)}_3 m_Q^5 \bar Q Q + \tilde c^{(X)}_5 m_Q^3 \bar Q \,i\sigma
G\,Q + \tilde c^{(X)}_6 m_Q^2(\bar Q\Gamma q)(\bar q \Gamma Q)+
{\cal O}(m_Q)\eqno(3)$$
with $i\sigma G=i\gamma_{\mu}\gamma_\nu G_{\mu \nu}$, $G_{\mu \nu}$ being the
gluonic field strength tensor.
The coefficients $\tilde c^{(X)}_i$ are calculable dimensionless numbers that
depend on the mode $X$.
Once the matrix elements between the states $H_Q$ are formed the first
term on the right hand side of eq.(3) yields (among other things)
the usual free `spectator'
result; the second one constitutes a new type of `spectator' contribution;
the third term represents `non-spectator'
contributions due to W-exchange (`WA') and interference effects (`PI').

(2) The relevant matrix elements of the operators
$\bar QQ$, $\bar Q\,i\sigma G\,Q$ and
$(\bar Q\Gamma q)(\bar q \Gamma Q)$
have been evaluated so far. For $\bar Q Q$ one employs a
nonrelativistic
$1/m_Q$ expansion:
$$\bar QQ=v_{\mu}\bar Q \gamma _{\mu}Q +
\frac {1}{4m_Q^2}\bar Q\,i\sigma G\,Q-
\frac{1}{2m_Q^2}\bar Q\,(D^2-(v\cdot D)^2)\,Q + {\cal
O}(1/m_Q^3)\;\;.\eqno(4)$$
$D_\mu$ denotes here the covariant derivative and $v_{\mu}$ is the
4-velocity vector. One finds:
$$ \la H_Q|\bar Q\gamma _{\mu}Q|H_Q\ra = 1\cdot v_{\mu} \mbox{ \hspace{2em}
( if } \;\la H_Q(\vec{p}\:')|H_Q(\vec{p}\,)\ra = \frac{E}{M} \cdot
(2\pi)^3 \,\delta^3 (\vec{p}\:'-\vec{p}\,)\; \mbox{ )}
\eqno(5)$$
$$ \la P_Q|\bar Q \,i \sigma G\, Q|P_Q\ra\,
\simeq \,\frac{3}{2}(M^2(V_Q)-M^2(P_Q))
\eqno(6)$$
$$ \la \Lambda _Q|\bar Q \,i\sigma G \,Q|\Lambda _Q\ra  \simeq 0\;;
\; \; \Lambda _Q=\Lambda _b, \Lambda _c\;;\; \;
P_Q=B,D\;; \; \; V_Q=B^*, D^*. \eqno(7)$$
The operator $\bar Q(D^2-(vD)^2)Q$ describes the kinetic energy of
the heavy quark in the gluon background field; the masses of
$\Lambda_Q$, $P_Q$, $V_Q$ yield some information on its size.

Finally we employ the approximation of factorization
$\la P_Q|(\bar Q\Gamma q)(\bar q\Gamma Q)|P_Q\ra \simeq
\la P_Q|\bar Q\Gamma q|0\ra \la 0|\bar q \Gamma Q|P_Q\ra $ with the operators
normalized at the low hadronic scale.

\vglue 0.6cm
{\elevenbf\noindent 2. Qualitative Results}
\vglue 0.4cm
The decay widths of a
heavy flavor hadron $H_Q$ look as follows:
$$\Gamma (H_Q)=\Gamma (Q)\cdot [1+a_2^{(H_Q)}/m_Q^2 + a_3^{(H_Q)}/m_Q^3 +
{\cal O}(1/m_Q^4)]\eqno(8)$$
where $\Gamma (Q)$ is the free quark width with all purely perturbative effects
included.

\noindent {\elevenbf (a)} The free
spectator ansatz thus holds through terms of order $1/m_Q$.

\noindent {\elevenbf (b)} The first nonperturbative corrections enter on the
$1/m_Q^2$ level with
$$ a_2^{(\Lambda _b)} \neq a_2^{(B^-)}=a_2^{(B_d)} \simeq a_2^{(B_s)}
\eqno(9)$$
{\elevenbf (c)} `Conventional' non-spectator
effects -- WA and PI -- emerge on the $1/m_Q^3$ level:
$$a_3^{(B^-)}<a_3^{(B_d)}<a_3^{(\Lambda _b)}\eqno(10)$$
{\elevenbf (d)} In charm decays these
nonperturbative corrections exhibit similar pattern while being larger by
roughly factors of
$(m_b/m_c)^2 \sim 10$ and $(m_b/m_c)^3 \sim 30$, respectively.

\vglue 0.5cm
{\elevenbf \noindent 3. (Semi-)Quantitative Phenomenology}
\vglue 0.2cm
{\elevenit \noindent 3.1 Semileptonic Branching Ratios: Corrections $1/m_b^2$}
\vglue 0.1cm
The semileptonic and nonleptonic transition operators for $b \rightarrow c$
decays in the external
gluon field are given by:
$$\hat \Gamma _{SL} = F\,\{z_0 \bar bb -\frac {z_1}{m_b^2}
\bar b\, i\sigma  G\,b\},\; \;
\hat \Gamma _{NL} = F N_C\{A_0(z_0 \bar bb -\frac {z_1}{m_b^2}
\bar b\,i\sigma G\,b)-\frac {4A_2z_2}{m_b^2}
\bar b\,i\sigma G\,b\}\eqno(11)$$
with $z_0=1-8x+8x^3-x^4-12x^2log(x)$, $z_1=(1-x)^4$, $z_2=(1-x)^3$ and
$x=m_c^2/m_b^2$ describing the relative phase spaces,
$A_0=(c_+^2+c_-^2)/2+(c_+^2-c_-^2)/2N_C$ and
$A_2=(c_+^2-c_-^2)/2N_C$ denoting the QCD radiative corrections
($N_C$=number of colours) and $F=G_F^2m_b^5|V_{cb}|^2/\,192 \pi^3$.
Thus we find a reduction in the semileptonic branching
ratio:
$$\frac {\delta BR_{SL}(B)}{BR_{SL}(B)}\simeq 6\cdot
\frac {M^2_{B^*}-M^2_{B}}{m_b^2}\frac {A_2z_2}{A_0z_0}\cdot
BR_{NL}(B)\; <\; 0\eqno(12)$$
This reduction is of the order of 5\% in B decays;
$BR_{SL}(\Lambda _b)$ remains practically unaffected by nonperturbative
corrections on
the $1/m_Q^2$ level due to eq.(7).

In charm decays on the other hand one obtains a much larger reduction:
$$\frac {\delta BR_{SL}(D)}{BR_{SL}(D)}\sim - {\cal O}(50 \%)\;\;.\eqno(13)$$
It is worth noting that in charmed particles some of the
${\cal O} (m_c^{-3})$ corrections are numerically as large as the
${\cal O} (m_c^{-2})$ ones.

The last term in eq.(11) is produced by an interference between two different
color flow amplitudes (it represents a non-factorizable contribution). It is
worth noting that it is approximately saturated$^4$ by non-factorizable
contributions to a few exclusive amplitudes that were estimated in ref.$^5$
arising due to the same operator.
\vglue 0.2cm
{\elevenit \noindent 3.2 Lifetime Ratios}
\vglue 0.1cm
Small differences are predicted for the lifetimes of the different beauty
hadrons:
$$ 1.02\; \; \leq \; \; \tau (B^-)/\tau (B_d) \; \; \leq
1.08 \mbox{ \hspace{3em} for \hspace{3em}} 130MeV < \;f_B \;< \;250MeV\;\;\;;
\eqno(14)$$
these effects are mainly due to `PI' with `WA' being rather small$^{1,2}$.
A preliminary analysis shows that $\tau (\Lambda _b)$ and $\tau (B)$ could
differ by up to 10-15 \%.

Extrapolating these findings down to charm one obtains much bigger
effects: $\tau (D^+)/\tau (D^0)\sim 2$ mainly due to `PI' with `WA' providing
a 20\% contribution
at most.
An intriguing `roller coaster' story thus unfolds behind the semileptonic
branching ratios for D mesons: on the $m_Q^5$ and $m_Q^4$ levels one gets the
parton result $BR_{SL}(D)\simeq 15\%$; $1/m_Q^2$ corrections suppress it
roughly by a factor of two: $BR_{SL}(D)\simeq 8\%$; $1/m_Q^3$ effects
finally differentiate between $D^0$ and $D^+$ decays by suppressing
$\Gamma_{NL}(D^+)$ such that $BR_{SL}(D^+)\simeq 16\%$ holds while affecting
$BR_{SL}(D^0)$ only a little.
\vglue 0.2cm
{\elevenit \noindent 3.3 $b\rightarrow s\gamma$}
\vglue 0.1cm
The chromomagnetic dipole operator
affects also radiative $b \rightarrow s + \gamma$ decays:
$$ \hat{\Gamma}_\gamma \;\propto \;m_b^5\,\{\: \bar{b}\,b\: -
\:1/m_b^2\:\bar{b}\,i\sigma G \,b\:\}\;\;; \eqno(15)$$
yet the ratio
$BR(B\rightarrow \gamma +(S=-1))\;/\;BR(B\rightarrow l \nu X)$ remains
practically unchanged.
\vglue 0.2cm
{\elevenit \noindent 3.4 Lepton Spectra in Semileptonic Decays}
\vglue 0.1cm
Lepton spectra in semileptonic beauty and charm decays can be
treated$^6$ in an
analogous fashion where one relies on an expansion in $1/(p_Q-p_l)^2$ rather
than in $1/m_Q^2$ with $p_Q\:[p_l]$ denoting the momentum of the heavy quark
$Q\:$
[the lepton l]. In the limit of the vanishing quark mass in the final
state of
$b\rightarrow q l \nu$ and neglecting the gluon bremsstrahlung one
obtains$^7$:
$$\frac {d\Gamma }{dy}\propto y^2\,[\:\frac {3}{2}-y
+\,(\,\frac{5}{3}y+\frac {1}{3}\delta (1-y)+\frac{1}{6}(2y^2-y^3)
\,\delta '(1-y))\:\frac{K}{2}+
(\,2+\frac{5}{3}y-\frac {11}{6}\delta (1-y))\:\frac{G}{2}\:]\;\;\eqno(16)$$
where $y=2E_l/m_b$,
$\;K=1/m_b^2\cdot \la H_b|\bar b \vec{D}^2b|H_b\ra $ and
$\;G=1/2m_b^2\cdot \la H_b|\bar b\,i\sigma G\,b|H_b\ra $.
The $\delta$-functions (for $m_q=0$) reflect the blowup of the expansion near
the endpoint. It is important however that the integral over a
small region $\Delta y > \mu /m_b$ ($\mu$ is a hadronic scale)
around the endpoint is correctly reproduced by that of eq.(16).
\vglue 0.5cm
{\elevenbf \noindent  4. Outlook}
\vglue 0.2cm
Obviously our analysis on the lepton spectra has to be
finalized$^7$. Beyond that
there are two intriguing areas for further study, namely

-- the question of $SU(3)_{flavour}$ breaking, e.g. the relation between
$D^0$ and $D_s$ and between $B_d$ and $B_s$ decays;

-- the vast realm of heavy flavour baryon decays.
\vglue 0.5cm
{\elevenbf \noindent 5. Acknowledgements \hfil}
\vglue 0.4cm
A.V. is grateful for helpful discussions to L.McLerran and M.Voloshin.
This work was supported in part by the NSF under grant
number NSF-PHY 92 13313 and in part by DOE under grant number
DOE-ER-40105-1100.
\vglue 0.5cm
{\elevenbf\noindent 6. References \hfil}
\vglue 0.4cm


\begin{thebibliography}{9}
\bibitem{VS} M.Voloshin and M.Shifman, Sov.J.Nucl.Phys. {\elevenbf 41}
(1985) 120; Sov.Phys.-JETP {\elevenbf 64} (1986) 698.
\bibitem{BU} I.I.Bigi and N.G.Uraltsev, {\elevenit Phys.Lett.}
{\elevenbf B280} (1992) 271.
\bibitem{BUV} I.I.Bigi, N.G.Uraltsev and A.Vainshtein,
{\elevenit Phys.Lett.} {\elevenbf B293} (1992) 430.
\bibitem{BS} B.Blok and M.Shifman,
preprints NSF-ITP-92-103, -115.
\bibitem{BS1} B.Blok and M.Shifman, preprint NSF-ITP-92-76.
\bibitem{CGG} The first detailled discussion of semileptonic spectra was given
in: J.Chay, H.Georgi, B.Grinstein,
{\elevenit Phys.Lett.} {\elevenbf B247} (1990) 399.
\bibitem{BSUV} I.I.Bigi, M.Shifman, N.G.Uraltsev and A.Vainshtein,
in preparation.
\end{thebibliography}
\end{document}